# Co-loading of Doxorubicin and Iron Oxide Nanocubes in Polycaprolactone Fibers for Combining Magneto-Thermal and Chemotherapeutic Effects on Cancer Cells

Francesca Serio,[1§] Niccolò Silvestri[2§], Sahitya Kumar Avugadda[2], Giulia E. P. Nucci[2], Simone Nitti[2], Valentina Onesto[1], Federico Catalano[2], Eliana D'Amone[1], Giuseppe Gigli[1,3], Loretta L. del Mercato[1*] and Teresa Pellegrino[2*]

1. Institute of Nanotechnology, National Research Council (CNR-NANOTEC), c/o Campus Ecotekne, via Monteroni, 73100, Lecce, Italy.

2. Istituto Italiano di Tecnologia, Via Morego 30, 16163 Genova, Italy.

3. Department of Mathematics and Physics "Ennio De Giorgi", University of Salento, via Arnesano, 73100, Lecce, Italy

§: these authors have contributed equally to this work

*: corresponding authors

**Abstract**

Among the strategies to fight cancer, multi-therapeutic approaches are considered as a wise choice to put in place multiple weapons to suppress tumors. In this work, to combine chemotherapeutic effects to magnetic hyperthermia when using biocompatible scaffolds, we have established an electrospinning method to produce nanofibers of polycaprolactone loaded with magnetic nanoparticles as heat mediators to be selectively activated under alternating magnetic field and doxorubicin as a chemotherapeutic drug. Production of the fibers was investigated with iron oxide nanoparticles of peculiar cubic shape (at 15 and 23 nm in cube edges) as they provide benchmark heat performance under clinical magnetic hyperthermia conditions. With 23 nm nanocubes when included into the fibers, an arrangement in chains was obtained. This linear configuration of magnetic nanoparticles resemble that of the magnetosomes, produced by magnetotactic bacteria, and our magnetic fibers exhibited remarkable heating effects as the magnetosomes. Magnetic fiber scaffolds showed excellent biocompatibility on fibroblast cells when missing the chemotherapeutic agent and when not exposed to magnetic hyperthermia as shown by viability assays. On the contrary, the fibers containing both magnetic nanocubes and doxorubicin showed significant cytotoxic effects on cervical cancer cells following the exposure to magnetic hyperthermia. Notably, these tests were conducted at magnetic hyperthermia field conditions of clinical use. As here shown, on the doxorubicin sensitive cervical cancer cells, the combination of heat damage by magnetic hyperthermia with enhanced diffusion of doxorubicin at therapeutic temperature are responsible for a more effective oncotherapy.

**Keywords:** electrospinning, iron oxide nanocubes, doxorubicin, stimuli-responsive nanofibers, magnetic hyperthermia, heat-mediated drug release.

**Introduction**

Most drug therapies used to treat cancer are systemic.[1] The main drawbacks of non-specific drug therapy are poor bioavailability, high-dose requirements, adverse side effects, development of multiple drug resistance, and non-specific targeting.[2] One way to overcome the above-discussed limitations is to develop drug delivery systems that promote controlled and targeted release of the existing drugs to the desired sites of therapeutic action while reducing adverse side effects.[2] Nowadays, several controlled release systems such as nanoparticles (NPs)[3], nanocapsules[4], micellar systems[5–7], and conjugates[8,9] have been well developed by encapsulating therapeutic agents in biodegradable matrices, such as polyester polymers.[10–13] However, the development of new nano-based formulations, with externally triggered actuation release still remains a great challenge but also a great opportunity to achieve controlled therapeutic action. Electrospun nanofiber mats were tested as drug delivery platforms for cancer research both *in vitro*, as scaffolds providing physiologically relevant 3D cancer models mimicking *in vivo* tumor micro-environment,[14] and *in vivo* as patches for therapeutic delivery.[15] In this regard, nanofibrous membranes have been implanted in the tumor cavity after surgical resection of the primary solid tumor, to inhibit the risk of cancer relapse and to promote controlled and sustained release of drugs at the tumor site, improving their efficacy and reducing adverse effects.[15–19]

Moreover, electrospun fibrous mats have recently attracted great attention as transdermal drug delivery systems (TDDS) for clinical applications: they can be designed to be applied as patches onto the skin surface and to deliver the active drug across the skin into the systemic circulation. Thanks to their porous architecture and to their ability of easily encapsulating drugs within the fiber lumen, higher drug loading and entrapment efficiency are reached compared to delivery systems prepared by other methods.[20] This is particularly relevant in the case of drug systems for cancer treatment, which often suffer of low drug concentration at cancer sites, poor solubility and instability in the biological environment.[21]

However, despite the many advantages of biodegradable electrospun materials in tuning the fiber morphology, porosity, and composition, the burst release of encapsulated drug represents the main drawback .[11,22] Very recently, increased attention has been given to composite drug delivery fibers obtained by embedding nano- and microparticles, which serve to either improve the drug retention into fibers or to allow a sustained release.[13,15,23,24] In this framework, the electrospinning technique represents an ideal fabrication method since it enables the easy development of composite multifunctional fibers with diameters ranging from a few tens of nanometers to a few micrometers by using polymeric mixtures and colloidal solutions.[25–29] Among a number of nanoscale materials being investigated for adding new functions to electrospun fibers, iron oxide nanoparticles (IONPs) are very attractive owing to their intrinsic magnetic properties with a great potential in many biomedical applications, such as diagnosis, imaging and magnetic hyperthermia (MHT) treatment.[30–35] In cancer therapy magnetic NPs can serve as heat mediators when exposed to an alternating magnetic field (AMF) in the so called MHT treatment.[36,37] This represents a powerful approach for killing cancer cells if therapeutic temperatures (≈45°C) can be reached.[38–40] Few recent works focus on the fabrication of multifunctional electrospun nanofibers doped with magnetic nanomaterials and show their superior properties for tissue engineering applications as well as for cancer treatment.[41–43] For instance, Li *et al.* discovered superior osteogenesis *in vitro* and bone regenerative ability of magnetic nanofibrous scaffolds made out of polycaprolactone (PCL) incorporating magnetic NPs.[42] In a different approach, Sasikala *et al*. applied magnetic nanofibers, made out of poly (D,L-lactide-co-glycolide), IONPs and the drug bortezomib, for *in vivo* MHT and drug release triggered by acidic tumor environment.[44] In a study by Kim and colleagues[24], a synergistic anticancer treatment was achieved by applying an AMF to electrospun NIPAAm-based nanofibers encapsulating doxorubicin (DOXO) and magnetic NPs. However, due to poor electrospinnability and biocompatibility of NIPAAm, its blending with other polymers or its chemical crosslinking by using thermal- or photocrosslinkable co-monomers is required.

In this work, we explored biodegradable and biocompatible PCL electrospun fibers co-loaded with magnetic iron oxide nanocubes (IONCs), having outstanding heat properties, to enable their activation under MHT conditions of clinical use and including within the lumen of the polymer fiber mats also the anticancer drug DOXO for combined hyperthermia with heat-enhanced drug release.

## Materials and methods

### Materials

Iron (III) acetylacetonate (Fe(acac)$_3$, 99%) and decanoic acid (DA, 99%) were purchased from Acros. Dibenzylether (DBE, 98%), acetone (99.5%), chloroform (99.8%), methanol (99.8%), isopropanol (99.8%), polycaprolactone (PCL, MW 80000 g mol$^{-1}$) and doxorubicin hydrochloride (98.0-102.0%) were purchased from Sigma Aldrich. Squalane (SQ, 98%) was purchased from Alfa Aesar. Milli-Q water filtered through 0.22 µm pore size hydrophilic filters (18.2 MΩ-cm) was supplied by a Milli-Q® integral water purification system. NIH/3T3 (CRL-1658), HeLa-WT (CCL-2) and MCF7 (HTB-22) cell lines were purchased from ATCC, dulbecco's modified eagle's medium (DMEM), RPMI 1640 cell culture media (RPMI), fetal bovine serum (FBS), penicillin streptomycin (PS), paraformaldehyde, triton X-100, dimethyl sulfoxide (DMSO), Phalloidin FITC-labelled, DAPI, MTT assay kit and glutamine were purchased from Sigma Aldrich.

### Magnetic iron oxide nanocubes synthesis

IONCs with edge length of 15 nm ± 1 nm and of 23 nm ± 5 nm were prepared by following a recently developed one pot synthesis approach as described previously by some of us.[45] Briefly, 0.353 g (1 mmol) of iron (III) acetylacetonate, with 0.69 g (4 mmols) of decanoic acid and 3 mL of dibenzyl ether (DBE) were dissolved in 22 mL of squalane (SQ). After degassing the solution for 120 minutes at 65°C, the mixture was heated up to 200 °C (3 °C/min) and kept at this temperature for 2.5 hours. The temperature was then increased at a heating rate of 7 °C/min up to 310 °C and maintained at this value for 1 hour. After cooling down to room temperature, the IONCs were then collected by precipitation with acetone followed by centrifugation at 8500 rpm. Finally, the supernatant was discarded and the black precipitate was washed twice in chloroform, collected and dispersed again in chloroform. The different IONCs sizes were obtained by varying the ratio between squalane and DBE solvents: ratio of 18:7 v/v in the case of IONCs 23 nm-sized and 23:2 v/v for IONCs 15 nm-sized synthesis. The resulted IONCs size dispersions was estimated using transmission elecrtron microscopy (TEM, JEOL JEM-1011, an acceleration voltage of 100 kV).

### Electrospinning fabrication of magnetic and DOXO-loaded magnetic fibers

The samples of IONCs dispersed in chloroform were transferred in an ultrasound bath at 40°C for 20 minutes prior to the preparation of the composite solutions with the aim to well disperse the nanocubes in suspension and reduce IONCs clustering. Then, bulk PCL in at a concentration of 12% (w/v) was dissolved in the chloroform containing the IONCs dispersion (5.46 mg.mL$^{-1}$). Next, methanol was added as a second solvent at a ratio methanol to chloroform of 70:30 v/v, with the aim to improve the electro-spinnability of the solutions.[46] For producing drug-loaded magnetic fibers, methanol with doxorubicin hydrochloride (0.133 mg mL$^{-1}$) was added. Given the high solubility of DOXO and IONC in the PCL polymeric solution, which form a one single phase and homogenous solution, we hypothesize that 10 µg is the maximum amount of drug loaded per fiber sample assuming no loss during the preparation. 0.355 mg in Fe is the amount of IONCs as measured by elemental analysis per fiber sample. The resulting suspensions were vigorously mixed and then stirred overnight. Immediately before the electrospinning process, each solution was transferred in a 1 mL syringe. The spinning process was performed by a 21G stainless steel needle and a syringe pump (E-fiber, SKE Advanced Therapies) with a feeding rate of 0.5 mL.h$^{-1}$, and by applying a positive high-voltage potential of 21 kV between the metallic needle and the collecting equipment of aluminum foil. The distance from the tip of the needle to the collector was adjusted to 15 cm. The air relative humidity and temperature during the electrospinning process were about 40% and 20°C, respectively. As controls, magnetic PCL fibers without DOXO, PCL fibers loaded only with DOXO, and neat PCL fibers, without either IONCs or DOXO, were fabricated as reference materials by using identical set-up parameters.

**Morphological characterization of electrospun fibers**

The morphology of the composite fibers was studied by scanning electron microscopy (SEM) analyses (MERLIN Zeiss) operating at an accelerating voltage of 5 kV. The samples were metallized with a thin layer of gold by using a sputter coater (Safematic CCU-010 LV vacuum coating) before SEM imaging. SEM micrographs were digitized and analyzed by ImageJ software to determine the average fiber diameters and distribution.[47] The average fiber diameters were calculated by analyzing a total of at least 100 fibers for each sample.

TEM analyses were carried out on a JEOL JEM-1011 instrument, using an acceleration voltage of 100 kV. To image the samples by TEM, the fiber mats were electrospun directly on TEM grids previously sticked onto the collecting screen by using bioadhesive carbon tape. The electrospinnig parameters described in section related to the magnetic scaffold fabrication were applied.

Confocal laser scanning microscopy (CLSM) was performed using a Leica SP8 microscope. DOXO-loaded fibers were imaged by using $\lambda_{exc}$: 514 nm and $\lambda_{em}$: 570-690 nm. The pinhole used for the imaging was 1 Airy unit. The emission was collected using 63x objective (HC PL APO CS2 63x/1.40 OIL) and the detector was PMT (PMT3, 570-690 nm).

**Heating performance characterization**

The hyperthermia efficiency of scaffolds was studied using a commercially available AMF calorimetric set up (DM100 series, nanoScale Biomagnetic Corp, available at Italian institute of technology, Genova). For analysis, a peel of fiber (one single preparation) was immersed in 200 µL of milli-Q water in a 0.3 mL glass vial and exposed to a field amplitude (H) of 30 kA.m$^{-1}$ and a frequency ($f$) of 110 kHz, for 30 min ( the H*$f$ product lies within the clinically and biologically acceptable limit, 5 x 10$^9$ Am$^{-1}$ s$^{-1}$).[48] The temperature rise in the glass vial was probed using a fiber optic probe (Fluoroptic® Temperature Probe, Model No. STF-2M, LUMASENSE™ TECHNOLOGIES), and corresponding temperature versus time (in seconds) curves were recorded. In order to evaluate the effect of field strength on heating abilities, at a fixed $f$ of 110 kHz, the PCL scaffold loaded with 23 nm IONCs (PCL-23_IONCs) fibers were exposed to different fields (16, 24, and 30 kA.m$^{-1}$). Similarly, the heating profile of PCL PCL-23_ IONCs fibers co-loaded with DOXO was tested in 200 µL phenol-free RPMI media at 110 kHz and 30 kA.m$^{-1}$.

**Cell Cultures**

Mouse embryonic fibroblast cell line (NIH 3T3 cells), DOXO-sensitive HeLa-WT cervical cancer cells and the DOXO-resistant MCF7 breast cancer cells were cultured in Dulbecco's Modified Medium (DMEM) supplemented with 10% Fetal Bovine Serum (FBS), 1% Penicillin Streptomycin (PS), 1% Glutamine and cultured at 37°C, 5% CO$_2$ and 95% humidity. Cells were split every 3-4 days, before reaching 75% confluence.

**Cell adhesion on composite fiber matrixes**

To this aim, mouse embryonic fibroblast cell line (NIH 3T3) at a density of $5 \times 10^4$ cells were grown directly on 2.2 cm diameter PCL-23_IONCs fiber matrix. Before cell seeding, scaffolds were sterilized with UV radiation from both sides for 1 hour each. All samples were assayed in triplicate and each sample was incubated in DMEM media supplemented with 10% of fetal bovine serum at 37 °C in a 5% $CO_2$ incubator (Thermo scientific Series 8000DH).

In order to investigate the cell morphology, the cytoskeleton and nuclei were stained after 7 days of cell culture. For CLSM imaging analyses, cells were fixed with 3.8% paraformaldehyde and treated with 0.1% Triton X-100 to permeabilize the cells membrane. Cytoskeleton filaments were stained with Phalloidin FITC-labelled 25 μg/ml, while nuclei were stained with DAPI 6 μg/ml. After staining, images were acquired by using a CLSM (Sp8 Leica). The same experiment was carried on neat PCL fibers used as control. Additionally, cell proliferation on PCL-23_IONCs fibers and neat PCL fibers was assessed by using the thiazolyl blue tetrazolium blue (MTT) assay after 1 and 7 days from cell seeding. Cells were incubated for 1 hour with MTT reagent at 37 °C and then were treated with DMSO solvent for 10 minutes at room temperature to dissolve formed blue Fformazan crystals. The DMSO solutions absorbance was measured at an optical density at 570 nm with a spectrophotometer (Cary UV-Vis 300).

***In vitro* cell viability assay of scaffolds with and without magnetic hyperthermia exposure**

For the evaluation of the chemotherapeutic effects combined with the MHT, different conditions were tested and are divided into four subgroups according to the treatment they underwent: 1) The control groups: include two samples: HeLa-WT cells exposed to neat PCL scaffold (PCL) and the . HeLa-WT cells not treated (CTRL) but kept at room temperature for 90 minutes, the same duration of a MHT treatment, before plating them back in culture. 2) The DOXO group includes three samples: PCL scaffold loaded with 10 μg of DOXO (PCL-DOXO), 10 μg of free DOXO added into the cells media (Free_DOXO), and PCL scaffold and free DOXO added into the cells media (10 μg) (PCL+Free_DOXO). In all these cases, no MHT was applied. 3) The IONCs group is composed of two samples: PCL scaffold loaded with IONCs without any MH treatment (PCL-23_IONCs) and PCL scaffold loaded with IONCs applying MHT (PCL-23_IONCs-MHT). 4) The combined treatment groups is composed of PCL scaffolds loaded with DOXO and: not exposed to MHT (PCL-23_IONCs-DOXO) or PCL scaffolds loaded with DOXO and exposed to MHT (PCL-23_IONCs-DOXO-MHT). For the experiment, the HeLa-WT and the MCF7 cell lines cultured in T-75 flasks were used. For each sample, once cells reached confluence, the old medium was substituted with a fresh one, and the scaffolds were added directly into the flask and kept in incubation for 24h at 37°C. Then, scaffolds were removed and placed in a glass vial (2 mL). The medium was preserved at 37 °C in a cell culture incubator. Meantime, cells were detached using 3 mL of Trypsin EDTA and centrifuged at 1000 rpm for 5 minutes. The pellet of cells (≈ 8 million cells per group), was resuspended in 200µl of fresh medium and added to the vial containing the scaffold. The MHT samples were exposed to three cycles of AMF, 30 minutes each, at a fixed frequency of 110 kHz and at a fixed temperature of 45°C (the field conditions varied from 24 to 30 kA/m and were adapted automatically by the AMF applicator to reach and maintain the target temperature). The rest of the samples were left at room temperature for 90 minutes. Later, the scaffolds were discarded and cells were seeded again in a 24 well-plate using the medium previously saved and stored. The viability was evaluated by measuring the cell metabolic activity through MTT assay at 24h, 48h, and 72h after the treatment. The experiment done on the MCF7 (DOXO-resistant) was also repeated using two scaffolds per flask (2 x PCL-23_IONCs-DOXO-MHT) to increase the available amount of DOXO.

**Results and discussions**

***Preparation and morphological characterization of the PCL magnetic fibers and*** ~~***doxorubicin***~~ ***drug loading***

The electrospinning technique has been employed for fabricating drug-loaded magnetic nanofibers. The spinning process was performed by using a PCL colloidal solution obtained by the dissolution of PCL polymer in a binary solvent system made by a 30 volume part of chloroform containing the IONCs dispersion and 70 volume part of methanol in which DOXO was dissolved. By applying an electric field, between the droplet of the solution loaded in the syringe at the needle tip and the collector plate, the solid nanofibers were deposited on the collector (schematic illustration of the electrospinning setup is available in Figure 1).

In order to reach the best magnetic heat performance, two different cube edge lengths of IONCs were tested: IONCs of 15 ± 1 nm (15_IONCs), and 23 ± 5 nm (23_IONCs), respectively, which generated the following two groups of DOXO-loaded magnetic fibers: PCL-15_IONCs-DOXO for fibers including 15 nm nanocubes and DOXO and PCL-23_IONCs-DOXO, for fibers including 23 nm nanocubes and DOXO.

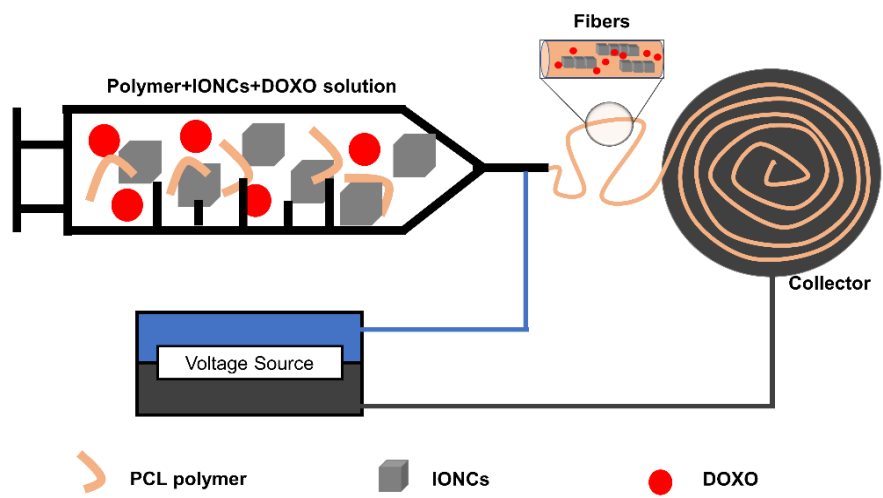

**Figure 1.** Schematic illustration of fabricating of magnetic fibers loaded with iron oxide nanocubes (IONCs) and DOXO, produced using an electrospinning setup.

SEM analyses, along with the calculation of average fiber diameters (Figure 2), revealed morphological differences between the different typologies of electrospun nanofiber matrixes. For instance, neat fibers (control samples) showed an uneven morphology (Figure 2a), while PCL-15_IONCs-DOXO and PCL-23_IONCs-DOXO composite fibers displayed uniform average diameter distributions (Figure 2b and c), with little change in fiber diameters, in accordance with the representative fitted Gaussian curves. Notably, with respect to the diameter distribution (1 ± 0.3 µm) of the neat PCL fibers (Figure 2d), it can be seen how after the addition of 15_IONCs (Figure 2e) structures became more regular and homogeneous with a maximum distribution peak around 1 ± 0.2 µm, while after the addition 23_IONCs (Figure 2f) the maximum distribution peak was around 0.7 ± 0.3 µm.

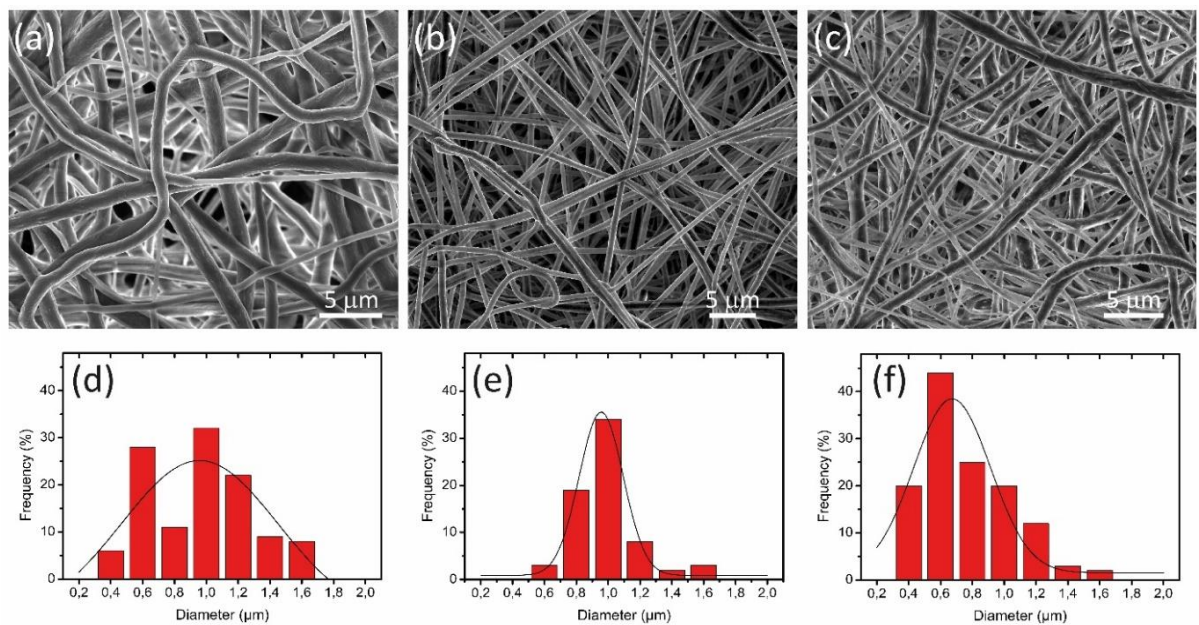

**Figure 2. SEM micrographs and fiber diameter distributions related to PCL samples.** SEM micrographs of (**a**) neat PCL fibers, (**b**) PCL-15_IONCs-DOXO fibers and (**c**) PCL-23_IONCs-DOXO fibers. The pictures show randomly distributed microfibers of the samples. Scale bars = 5 μm. (**d-f**) Graph illustration of average diameter distribution in (**d**) the neat PCL fibers, (**e**) PCL-15_IONCs-DOXO fibers, and (**f**) PCL-23_IONCs-DOXO fibers. The superimposed continuous lines are the best fits for Gaussian curves.

Furthermore, TEM analyses performed to visualize the arrangement of the 15_IONCs and the 23_IONCs within the polymer fibers, underlined a different distribution of the nanocubes into the fiber matrixes depending on the size of IONCs used (Figure 3).

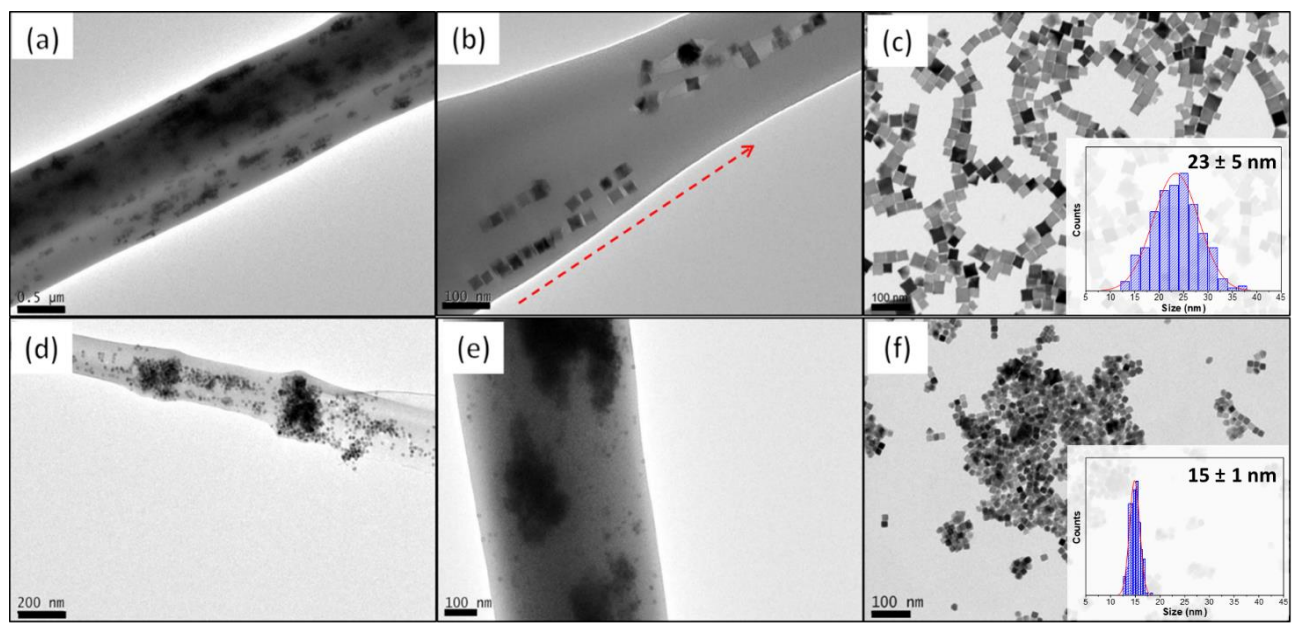

**Figure 3. TEM images of the DOXO-loaded magnetic fibers.** (**a** - **b**)-Images of PCL-23_IONCs-DOXO

fibers made with 23_IONCs. In (**b**) the spontaneous alignment (red arrow) of 23_IONCs along the fiber axis can be appreciated. (**c**) Free 23_IONCs (inset shows the average diameter distribution of nanocubes). (**d - e**) PCL-15_IONCs-DOXO fibers made with 15_IONCs. (**f**) Free 15_IONCs. Insets in panel c and f indicate the average size distribution of nanocubes and determined by TEM analysis).

In particular, a uniaxial alignment of 23_IONCs with respect to the fiber length was appreciated (Figure 3b). This strict control of the nanomaterial orientation is provided by the electrospinning process and, specifically, it may be due to an early alignment by the flow in the syringe needle followed by the jet stretching and thinning.[49,50] Such nanocubes alignment effect was observed only for 23_IONCs. It is worth noting, that magnetic nanoparticles may form chains in solution because of the magnetic coupling effects between the NPs, due to the strength of the particle-particle interactions.[51] This phenomenon is strongly size dependent and more prominent in magnetic NPs above the size of 20 nm.[52] Instead, for fully superparamagnetic nanocubes like the 15 nm IONCs, the inter-particle coupling was negligible and they appeared more clustered in the core of the fibers with a random distribution.

The presence of DOXO within magnetic PCL fibers (both in PCL-15_IONCs-DOXO and PCL-23_IONCs-DOXO) was assessed under CLSM at an excitation wavelength of 514 nm. The PCL fibers loaded only with DOXO (PCL-DOXO, without IONCs), and the neat PCL fibers (no DOXO and IONCs) were also imaged for comparison purpose, as shown in Figure 4.

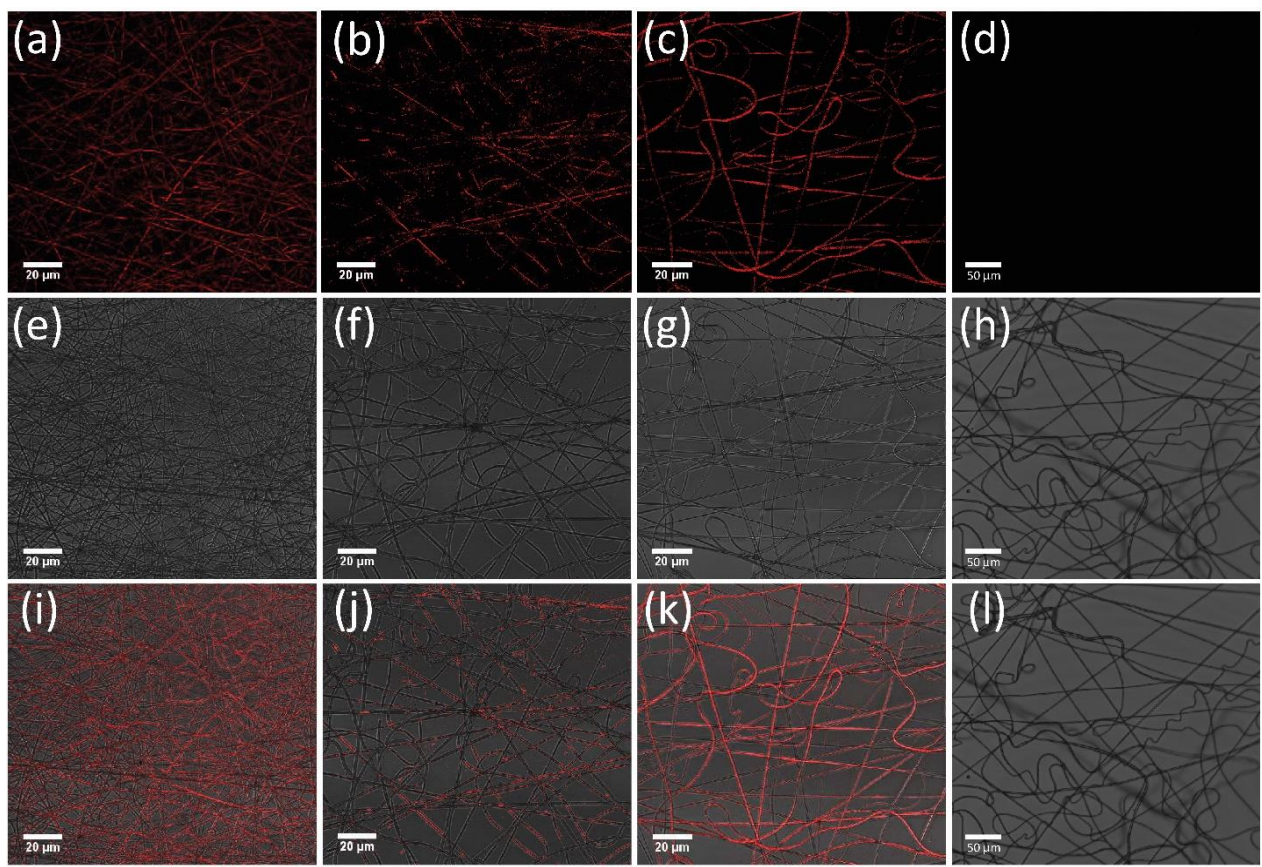

**Figure 4. Confocal images of PCL magnetic fibers co-loaded with DOXO.** CLSM images of (**a**) PCL-23_IONCs-DOXO, (**b**) PCL-15_IONCs-DOXO and (**c**) PCL-DOXO and (**d**) neat PCL fibers. Figures (**e-h**) show the corresponding bright field images of each sample shown in a-d. Figures (**i-l**) show merged images of fluorescent and bright field channels. ($\lambda_{exc}$: 514 nm; $\lambda_{em}$: 570-690 nm).

As expected, the inherited red fluorescence signal, owing to DOXO incorporation, was unambiguously detected in PCL-23_IONCs-DOXO, PCL-15_IONCs-DOXO, and PCL-DOXO (Figures 4 a, b and c) with respect to the neat PCL fibers, where no fluorescence was detected (Figures 4 d, h). Similarly, to the control sample PCL-DOXO (Figure 4c), the drug- fluorescence in PCL-23_IONCs-DOXO (Figure 4a) was found to be uniform thus indicating a homogeneous distribution of DOXO within the fibers. On the contrary, the distribution of DOXO in PCL-15_IONCs-DOXO resulted less homogeneous (Figure 4b), some fluorescent clots appeared within the fibers, likely related to the aggregation of 15 nm IONCs into the fibers as indicated by TEM analyses (Figures 3 d, e).

The quantification of DOXO release from electrospun fiber mats is challenging because the drug release process is influenced by several factors, including the properties and the structure of the fiber mats. These aspects make difficult the standardization of the methods for the analysis of drug release from fiber mats. The most used methods are based on collection of samples at different time points and analyses by ultraviolet-visible (UV-VIS) photoluminescent (PL) spectrophotometry or high performance liquid chromatography (HPLC).[53] In our case the DOXO release quantification after MHT treatment was not straightforward and the attempt to measure by UV-VIS and photo luminescent (PL) spectrophotometer methods the DOXO amount released by the scaffold under MHT treatment was unfortunately not successful (Figure S1). Indeed, on the PCL-23_IONCs-DOXO scaffold immersed in 200 µL phenol-free RPMI media after MHT (3 cycles, 30 minutes for each cycle at 110 kHz and 30 kA/m), and after having removed the scaffold, the UV-VIS and PL (excitation wavelength at 480 nm) spectra of the RPMI media was recorded. Likely due to the low DOXO loaded into the scaffolds (nominal amount only 10 µg/scaffold) and to the low DOXO released after the three cycles of MHT, the UV-VIS and the PL spectra did not show any spectral features that could be exploited for the quantification of the DOXO release upon MHT (Figure S1).

*Heating efficiency of the PCL magnetic fibers under MHT*

The MHT performance of electrospun magnetic fibers were tested by dipping the as-synthesized fibers in water and measuring with an optical probe the temperature increases of water overtime upon applying an external AMF (Figure 5a). On exposing these fibers for 30 minutes to AMF (110 kHz and 30 kA/m), it is evident that PCL-23_IONCs fibers achieved a maximum temperature of 38.5 °C with respect to 30.5 °C reached by PCL-15_IONCs under the same AMFs conditions. To note that the same amount of nanocubes (in iron amount), was used for the preparation of the PCL-15_IONCs and PCL-23_IONCs fibers and the AMFs conditions chosen for these tests were the ones clinically used (Figure 5b). The heat difference of 20.5 °C for PCL-23_IONCs *versus* the 12.5 °C for PCL-15_IONCs (Figure 5b, inset), can be attributed to the highest specific adsorption rate values of the 23 nm particles than that of the sample at 15 nm cubes, as previously reported.[45] In addition, the alignment of the nanocubes within the fibers (Figure 3b) may also favorably affect the heating performance of the magnetic fibers. Indeed, previous works have shown, how the different chain-aligned configurations (*i.e.* magnetosome chains in magnetotactic bacteria, colloidally prepared chains) had enhanced the overall heating properties of magnetic NPs dispersion. The magnetic NPs in uniaxial assembly, indeed, could interlock their individual magnetic moments along the direction of chain-length under applied magnetic fields, resulting in large dynamic hysteresis loop than single particle and increase heating performance.[52,54–58] On the contrary, nanocubes of 15 nm in size in PCL-15_IONCs (Figures 3c,d), despite reaching lower temperature under the same AMF conditions due to their smaller size than the 23 nm nanocubes, appeared aggregated, forming random and inhomogeneous clusters within the fibers. This may also contribute to demagnetization effects, which do not favor their heat abilities in MHT. For this reason, for most of the study with cells, we have selected the fibers containing the 23_IONCs. Further calorimetric characterization of PCL-23_IONCs, at a frequency of 110 kHz, the temperature versus time curve fibers were recorded at different magnetic field amplitudes and as expected, the fibers were heating faster and reached higher temperatures when exposed to higher magnetic field intensities. Indeed, the temperature increase within the first few thousand seconds was negligible at 16 kA/m, reached 27.7 °C at 24 kA/m, and the 38.5 °C at 30 kA/m (Figure 5c).

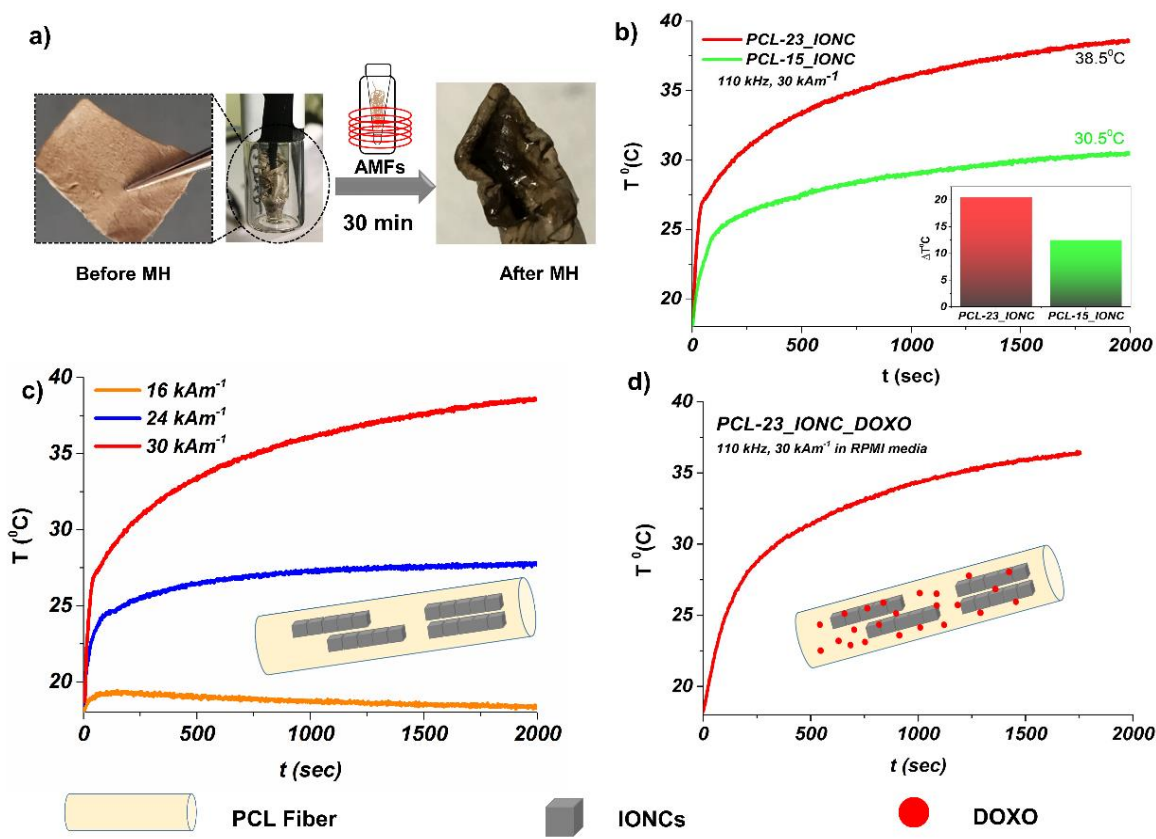

**Figure 5. Heating efficiency of fibers under alternating magnetic fields**. (**a**) Magnetic fiber matrix peeled off from the collector and placed in an Eppendorf with Milli-Q water in which a thermoprobe was set to record the temperature increase during the AMF application at 110 kHz. (**b**) Comparison of the Temperature (T °C) versus time curves for PCL-23_ IONCs and PCL-15_IONCs in the water at a frequency of 110 kHz and magnetic field intensity of 30 kA/m. Inset shows the heat difference (ΔT°C) achieved by each scaffold after 30 minutes of exposure to the AMF. (**c**) T versus time curves of PCL-23_IONCs fibers at a fixed frequency of 110 kHz and field amplitude intensities varied from 16 kA.m$^{-1}$ to 30 kA/m. (**d**) Heating profile of PCL-23_IONCs loaded with DOXO, immersed in phenol-free RPMI media achieved a maximum of 36.5 °C with an obtainable ΔT of 18.5 °C.

It is worth mentioning, the magnetic fiber mats at the end of the MHT heating cycle appeared to undergo some morphological changes but, as assessed by elemental analysis, no release of nanocubes was measured in the cell media (Figure S2). The same mat just before the exposure (Figure 5a), right after the AMFs treatment, was crumpled upon itself and a brownish colour appeared most likely because of the chemical change of the nanocomposite fibers which melt on the IONCs surrounding (Figure. 5a, right image). Indeed, as shown by others the surface temperature of magnetic nanoparticles under AMF is likely much higher than 45°C that is the temperature measured in the bulk solution.[59–61] A close look at the magnetic fibers by scanning electron microscopy also indicate a significant structural modification of the fibers upon exposed to MHT (Figure S3). It is likely that since PCL has the melting point around 60°C the 90 minutes exposure to heat under MHT, has accounted for a partial melting process of the fibers although the macroscopic temperature measured in solution was of 45°C (Figure S3).

Moreover, when applying AMFs to PCL-23_IONCs fiber, in absence of any solution/media, a rapid shrinking and change in color of the scaffold was observed (Figure S4 inset). Also, when the optical thermal probe was touching directly the scaffold, a drastic increase of temperature difference up to 38 °C was reached, within few seconds (≈ 14 sec) (Figure S4). This steeper temperature change for magnetic fibers measured directly at the fiber's surface with the optical probe is likely due to the quicker heat transfer in solution free-state. Instead, in water or in RPMI media, the slower temperature increase is likely attributed to the specific heat capacity of water and media which affect the heat transfer from the fibers to the surrounding media (Figure S5). For the sample PCL-23_IONCs-DOXO, the co-loading of DOXO in the fibers with IONCs does not affect the heating properties of the nanocubes. Indeed, as shown in Figure 5d, the T *versus* time heat curve at AMFs (110 kHz and 30 kA/m) for the PCL-23_IONCs-DOXO immersed in 200 µL of phenol-free RPMI media, exhibited a comparable profile with that of PCL-23_IONCs fibers in water and not bearing DOXO (ΔT of 18.5 °C reached by PCL-23_IONCs-DOXO is close to 20.5 °C of PCL-23_IONCs).

*Biocompatibility of the magnetic PCL fibers*

To assess the intrinsic scaffold's biocompatibility before their exposure to external stimuli, such as AMF, the neat PCL fibers and the PCL fibers loaded only with IONCs (PCL-23_IONCs) were exposed to 3T3 cells (3-day transfer, inoculum 3×10[5]) by seeding the 3T3 cells on the fibers. The cell's morphology was studied after 7 days of cell culture. The 3T3 cells imaged by CLSM, following labeling of cytoskeleton filaments with phalloidin FITC and nuclei labeling with DAPI, were found to nicely adhere on the fibers with no sign of cell sufferance. Indeed, there were no morphological differences in cells seeded on neat PCL (Figure 6a and c) and PCL-23_IONCs fibers (Figure 6b and d). In both cases at 7 days, cells formed continuous cell layers growing on the fibers mats. The obtained data indicate that the PCL-23_IONCs fibers do not hinder cell adhesion and cell growth.

Additionally, cell proliferation on PCL-23_IONCs fibers was compared with that on neat PCL fibers. This viability test was assessed by using the cell metabolic activity MTT assay at 1 and 7 days and each absorbance at 570 nm of Formazan as a product of the metabolic activity, was measured with a spectrophotometer (Figure 6e). The absorbance detected at day 1 for PCL-23_IONCs fibers was the same value of those measured of neat PCL fibers (with no IONCs). At day 7, no significant difference in viability was observed between the two samples, confirming PCL-35_IONCs fibers as a biocompatible and non-toxic substrate for cell culture.

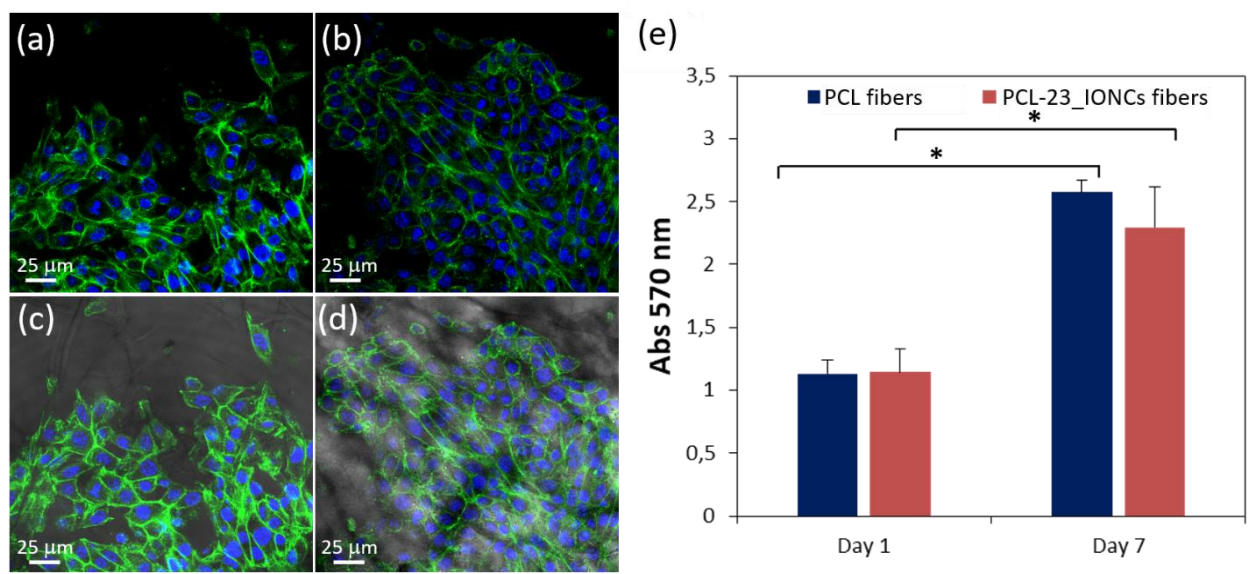

**Figure 6. Biocompatibility and cell proliferation of 3T3 cells on magnetic fibers.** CLSM images of 3T3 cells seeded on (**a**) neat PCL fibers and on (**b**) PCL-23_IONCs fibers. In (**c**) and in (**d**), the respective overlay images of bright field and fluorescence images are reported. Actin filaments are shown in green, nuclei are shown in blue. (**e**) Graph displaying the absorbance of formazan crystals dissolved in DMSO after the reaction between 3T3 cells and MTT reagent at 1 and 7 days from cell seeding. Results are expressed as (mean ± standard deviation). Bars show statistically significant differences (Number of replicates: N = 3; *P < 0.05; one-way ANOVA)

Published in Journal of Colloid and Interface Science, Volume 607, Part 1, February 2022, Pages 34-44, Doi: https://doi.org/10.1016/j.jcis.2021.08.153*In vitro viability assay on tumor cells: verifying the synergic action of DOXO release and MHT heat treatment*

Having established the biocompatibility of the magnetic fibers in absence of drug, we next investigated the combined efficacy of MHT and the chemotherapeutic effect of PCL-23_IONCs-DOXO scaffold loaded with DOXO to kill tumor cells.

A viability study was performed on the DOXO sensitive epithelial cervical cancer (cell line HeLa-WT). The samples were exposed to different materials (PCL- 23_IONCs scaffold loaded with DOXO or PCL-23_IONCs) and to the presence or absence of the MHT. The main control group consisted of Hela-WT cells, which were not undergoing any treatment (CTRL). Moreover, the PCL scaffold sample, which did not contain IONCs nor the DOXO, was also used as a reference control. More specifically, to evaluate the effects of heat damage, the cell exposed to sample PCL-23_IONCs fibers with MHT were directly compared to a cell sample treated with PCL-23_IONCs fibers but not exposed to the AMF. To evaluate the drug effects, PCL-23_IONCs scaffold loaded with DOXO, with or without exposure to MHT, was compared to additional two samples. The first one included PCL-DOXO (fibers loaded with only DOXO), which enable to evaluate the effect of the passive and non-specific DOXO release. The second one consisted of Free_DOXO, a cell sample treated with a dose of DOXO (10 µg) that is equivalent to the nominal loaded amount of DOXO within the fibers.

For the experimental scheme (Figure 7.I), the tumor Hela-WT cells were incubated with each of the fiber samples described above for 24h. Scaffolds were added directly into the medium while the cells were adhering on the bottom of the flask forming a two-dimensional cell layer. After the 24 h incubation period, the cells were detached and the cell pellet and the scaffold were transferred in the same glass vial (Figure 7.II). 200 ul of medium were left in the vial in order to avoid the dryness of the pellet. The culture media were saved and stored at 37°C to be used to re-plate the cell after the application of MHT. This was done to take into account also the passive release of DOXO and the possible degradation of the PCL scaffolds into the media.

The glass vials containing the cells and the PCL-23_IONCs or the cells and the PCL-23_IONCs-DOXO were exposed to three cycle of MHT (f=110 kHz, H=24-30 kA/m, 30 minutes each). These MHT samples were named respectively PCL-23_IONCs-MHT and PCL-23_IONCs-DOXO-MHT). In these experiments, the field intensity was adjusted automatically by the AMF applicator ensuring that the IONCs maintained the target therapeutic temperature of 45°C (Figure 7.III). All the other samples (CTRL, PCL, Free_DOXO, PCL+Free_DOXO, PCL-DOXO, PCL-23_IONCs and PCL-23_IONCs-DOXO) were handled in the same way, but, instead of exposing them to AMF, they were left at room temperature for the same duration time of the whole hyperthermia treatment (90 minutes). After the treatment, the cells were re-plated with the saved media (the same media used for the first 24 h incubation) and the scaffold were discarded. The effects of the single therapies and the combined effects were assessed through the application of MTT assay at 24h, 48h and 72h from the treatment (Figure 7.IV).

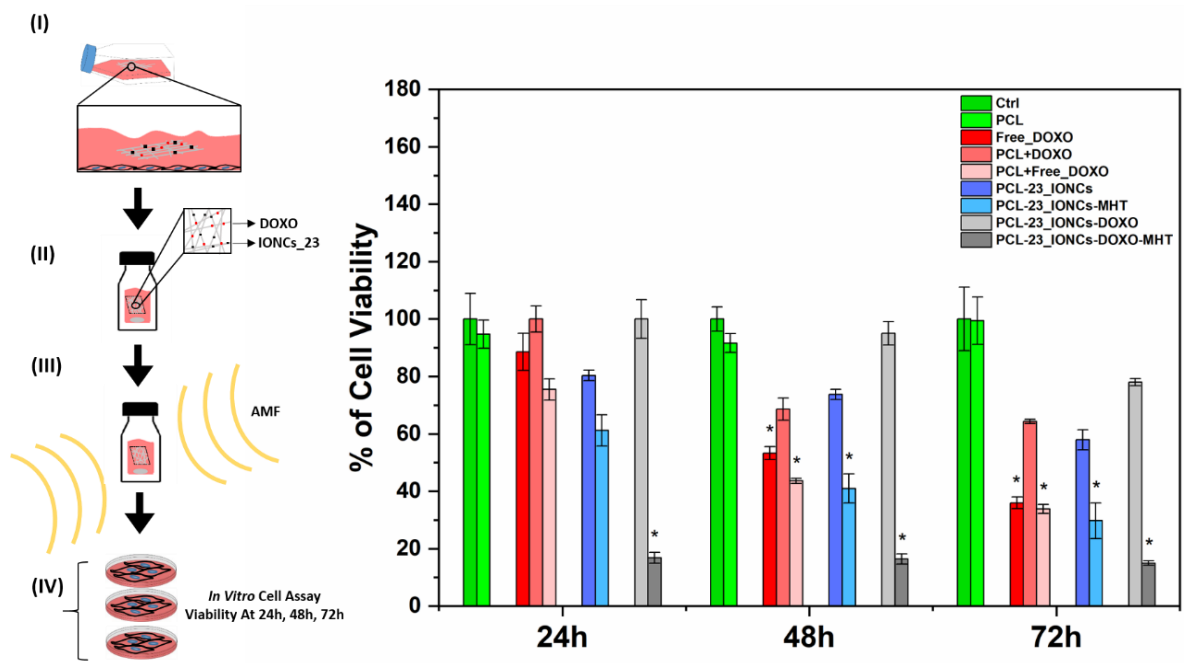

**Figure 7.** Cells viability experiment on HeLa cells exposed to PCL scaffold loaded with IONCs and DOXO after MHT alone and MHT combined with DOXO treatment, in comparison with the passive release of the DOXO. Schematic representation of the experiment: **(I)** In the first step, scaffolds were added to the cultures and the flasks were kept for 24h at 37°C. **(II)** After the incubation, cells were collected together with the scaffolds and placed in a glass tube. **(III)** Then, the PCL-23_IONCS-MHT and PCL-23_IONCs-DOXO-MHT samples were exposed to three cycles (30 minutes each) of MHT, in AMF (f= 110 kHz H= from 24 to 30 kA/m adjusted automatically by the AMF applicator to maintain the temperature: T=45°C). The other samples were left 90 minutes at room temperature. **(IV)** After the treatment step, PCL scaffolds were discarded and the cells were re-plated in a 24 well-plate. The viability was assessed by measuring the metabolic activity through MTT assay at 24h, 48h and 72h after the treatment. The samples are divided into 4 subgroups. The first one is the control group, CTRL sample just Hela cells (in green), and the PCL scaffold alone (in light green). The DOXO group (no MHT applied), cells were exposed to DOXO in order to check the passive release of the drug, includes three samples: the PCL scaffold loaded with DOXO only (PCL-DOXO) (in red), a sample of DOXO added in to the cells media at the same concentration loaded within the fibers, 10µg

(Free_DOXO) (in light red) and PCL scaffold only with free DOXO added at the same concentration loaded within the fibers (PCL+Free_DOXO) (in pink). The group of PCL scaffold loaded just with 23_IONCs. These samples are PCL scaffold loaded with IONCs without MHT applied (PCL-23_IONCs) (in blue) and applying MHT (PCL-23_IONCs--MHT) (in light blue). The last group: combined treatment, without MHT applied (PCL-23_IONCs-DOXO) (in light-grey) and applying MHT (PCL-23_IONCs-DOXO-MHT) (in dark-grey). Results are expressed as (mean ± standard deviation). Stars show statistically significant differences compared with control, CTRL (Number of replicates: N = 3; *p < 0.05; one-way ANOVA).

The viability results of the experiment show that the PCL-23_IONCs-DOXO-MHT sample, in which the heating treatment is combined with the drug, is the most cytotoxic against the tumor cell line used (Figure 7). In fact, the cells viability drastically decreased to 16% already after the first 24h from the MHT and reached 15% at 72h. The temperature increase driven by the 23_IONCs under AMF also promote the release of the DOXO into the media resulting in a combined toxicity action detected by the reduced viability of the cells. Indeed, the sample PCL-23_IONCs-MHT, which does not include DOXO is less effective than the sample PCL-23_IONCs-DOXO-MHT, provides cytotoxicity effects that are still significant on this tumor cell line: after 24h from the treatment, the percentage of cells alive was 61% and decreased over time, reaching 29% at 72h. This result suggest that the heat released by the IONCs in AMF causes cells damages and induces the apoptosis process that leads to a drop in cells viability over the time window studied. The viability of the cells incubated with PCL-23_IONCs sample, instead, decreased from around 80% at 24h to 60% after three days. Despite the value is lower than the other controls used, 60% is not to be considered a threshold for toxicity of the sample. The passive drug release (PCL-DOXO and PCL-23_IONCs-DOXO) was analyzed and compared to the exposure to the free drug (Free_DOXO and PCL+Free_DOXO). It is evident from the chart, that there is a reduction of cells proliferation below 60% in the cases where free Doxo was present, results that are coherent with the literature.[62] The samples with the drug loaded into the fibers (PCL-DOXO and PCL-23_IONCs-DOXO) show less toxicity with respect to the others samples in which free DOXO has been administered (Free_DOXO and PCL+Free_DOXO) indicating that a modulation of the DOXO release occurs whenever the drug is encapsulated within the fiber mat. Indeed, this is due to the PCL scaffold composition, that being biodegradable, releases a small percentage of the drug loaded which inevitably affects the kinetic drug release. However, this passive release is not effective, since at 72h there is still a vitality around 65%. Finally, the control group, CTRL and PCL samples, show 100% and 95% of cell viability along the whole experiment, highlighting the non-toxicity of the scaffold itself. Overall, although indirectly, these data suggest, the cytotoxic effects of DOXO drug release and at the same time, they highlight the importance to couple the heat-mediated drug release with the MHT heat damage for a more effective antitumoral therapy.

It is also worth to mention the results of cytotoxicity obtained when the same experiment was performed on a cell line resistant to DOXO drug, the MFC7 breast cancer cell line. In this case, for the most effective sample, the PCL-23_IONCs-DOXO-MHT, the viability decreased below 20% after the MHT in the first 24h (Supplementary Information, Figure S6), highlighting one more time the effectiveness of the thermal treatment. However, in the subsequent 24h, cells were able to regrow (viable cell percentage raised around 44%). The short-term damage caused by the MHT was not supported by the long-term damage caused by the drug, being the cells resistant to DOXO. To check whether the resistance of these cells was related to the drug concentration used (10 µg), the experiment was carried out by exposing the cells at doubling concentration of DOXO (20 µg), namely incubating the cells with two PCL-23_IONCs-DOXO-MHT scaffolds instead of one. Nonetheless, the viability of the cells fluctuated between 60% and 70% among the time span of the experiment, revealing that this treatment is not sufficiently efficient on DOXO resistant cells line (Supplementary Information, Figure S7). Further investigations are needed to find a proper therapy for this type of cells.

**Conclusions**

In this work, the design, fabrication and characterization of biocompatible fibers co-loaded with a chemotherapeutic agent, DOXO, and magnetic nanoparticles of cubic shape have been reported. Particularly, we here aimed at fabricating magnetic fibers having optimized MHT heat performance under clinically safe conditions.[63–65] The electrospinning of a homogenous phase solution containing 23 nm (or 15 nm) IONCs as magnetic nanoparticles, chosen for their MHT heat performance and the PCL polymer, guarantees the fabrication of magnetic fibers of about 0.5-1 µm diameter. The morphology characterization by TEM analyses showed the clear incorporation of the nanocubes within the fibers and highlighted also, the preferential alignment of nanocubes in small chains within the fiber length only for the 23 nm nanoparticles. Next, the co-loading of DOXO within the magnetic fibers, as confirmed by the bright PL DOXO signal under CLSM analyses, was made possible by simply adding the drug to the electrospinning solution containing the PCL polymer and the magnetic nanocubes. Additionally, the magnetic properties of PCL-23_IONCs-DOXO fibers have also shown the quick raise of temperature profile under clinically safe magnetic field conditions. Finally, the drug-free PCL-23_IONCs fibers showed high biocompatibility when used as substrates for growing epithelial 3T3 cells with no adverse effects on cell adhesion and viability in absence of MHT exposure. Instead, the electrospun magnetic PCL fibers mats co-loaded with DOXO show cytotoxic effects against a DOXO sensitive cell line, HeLa_WT cells especially in presence of the MHT treatment. Indeed, the heat caused by MHT together with the release of DOXO resulted in an increased mortality (cells viability of less than 20%) in the cell group treated with PCL-23_IONCs-DOXO and exposed to MHT treatment as compared to the single therapies (only non-specific drug release and only MHT with no drug loaded). Given the simple scalability of the electrospinning PCL-fiber and magnetic nanocubes production, the fabrication process results straightforward and inexpensive. Overall, these results suggest the potential use of these magnetic fibers as nanoplatform in hyperthermia cancer treatment combined with the delivery of cytotoxic agents. In the near future, the co-encapsulation of these MHT efficient nanocubes and DOXO within polymer fibers having a slower degradation profile[66] may also provide a prolonged drug toxic effect, facilitating the eradication of DOXO-resistant cancer cells in aggressive primary tumors. In perspective, we may also envision the exploitation of such scaffolds as implantable tool at solid tumors (*e.g.* Glioblastoma multiforme for which MHT is already applied in clinic)[67] that enables to provide a reservoir of efficient magnetic nanoparticles for MHT treatment while promoting the heat-boosted release of DOXO followed by local and long-term drug depot release from the scaffolds. This will require further characterize with the right *in vivo* tumor model.[68]

**Acknowledgements**

L.L.d.M. gratefully acknowledges support from the European Research Council (ERC) under the European Union's Horizon 2020 research and innovation program ERC Starting Grant "INTERCELLMED" (contract number 759959), the My First AIRC Grant (MFAG-2019, contract number 22902) and the "Tecnopolo per la medicina di precisione" (TecnoMed Puglia) Regione Puglia: DGR n.2117 of 21/11/2018, CUP: B84I18000540002. T.P. is grateful for the funding support to European Research Council (starting grant ICARO, contract number 678109) and to the AIRC association (AIRC–IG, contract number 14527).

## Graphical abstract

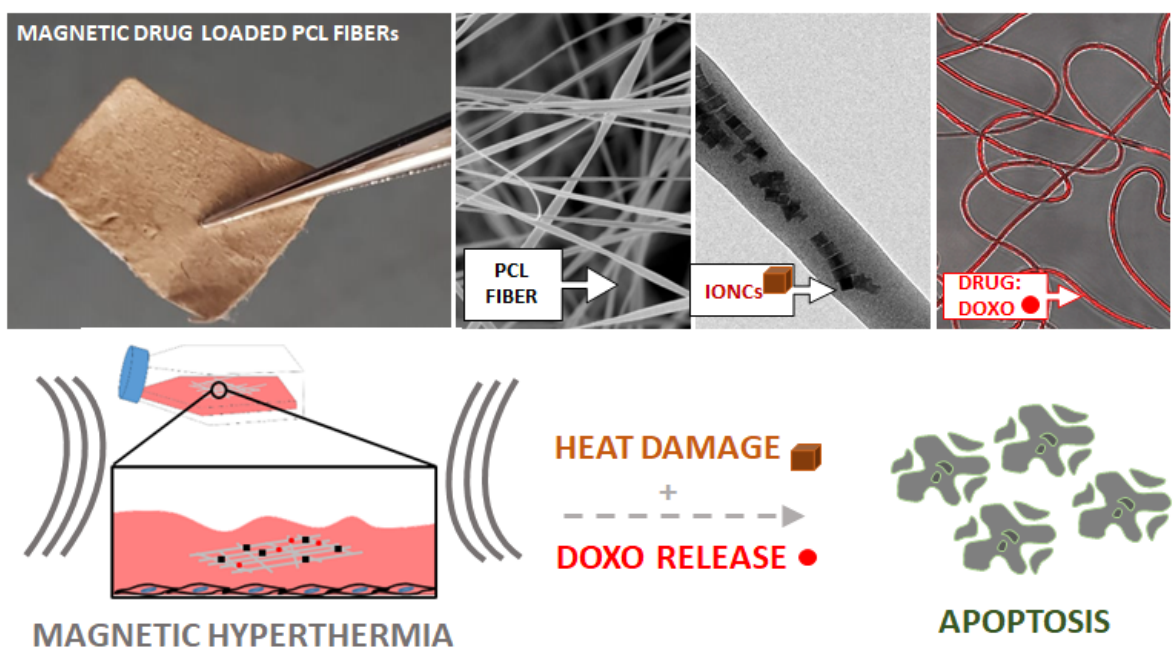

**Electronic supporting information**

# Co-loading of Doxorubicin and Iron Oxide Nanocubes in Polycaprolactone Fibers for Combining Magneto-Thermal and Chemotherapeutic Effects on Cancer Cells

Francesca Serio,[1,§] Niccolò Silvestri,[2,§] Sahitya Kumar Avugadda,[2] Giulia E. P. Nucci,[2] Simone Nitti,[2] Valentina Onesto,[1] Federico Catalano,[2] Eliana D'Amone,[1] Giuseppe Gigli,[1,3] Loretta L. del Mercato[1,*] and Teresa Pellegrino[2,*]

1. Institute of Nanotechnology, National Research Council (CNR-NANOTEC), c/o Campus Ecotekne, via Monteroni, 73100, Lecce, Italy.
2. Istituto Italiano di Tecnologia, Via Morego 30, 16163 Genova, Italy.
3. Department of Mathematics and Physics "Ennio De Giorgi", University of Salento, via Arnesano, 73100, Lecce, Italy

§: these authors have contributed equally to this work

*: corresponding authors

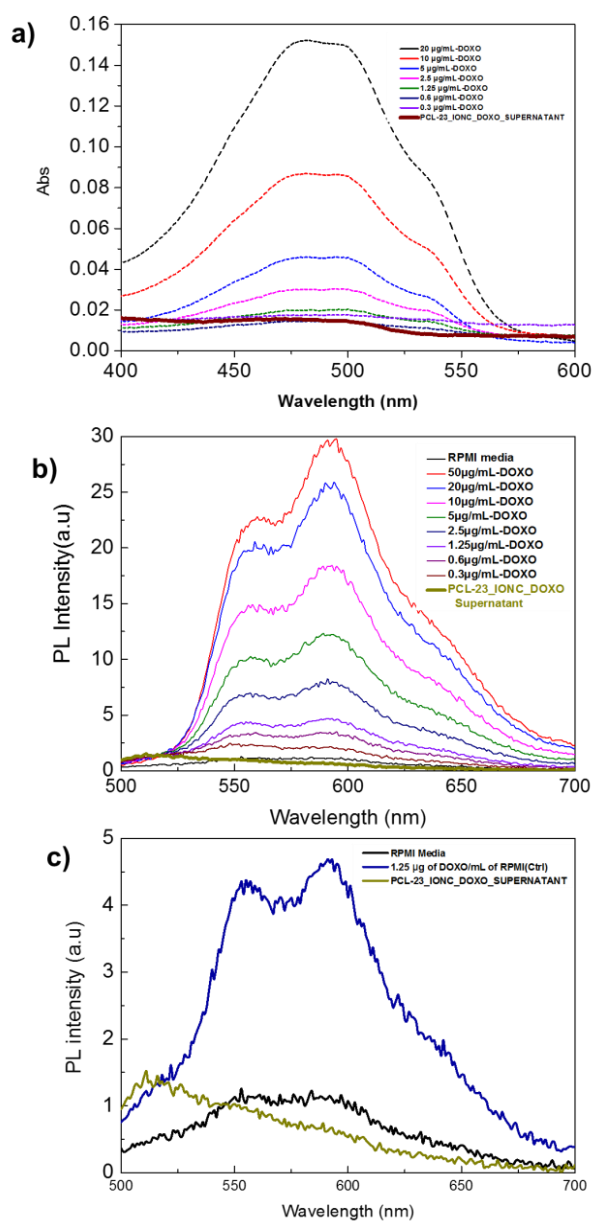

**Figure S1. UV-visible and PL spectra (wavelength of excitation 480 nm) of the PCL-23_IONCs-DOXO solution after exposure to MH treatment.** The PCL-23_IONCs-DOXO immersed in phenol-free RPMI media was exposed to MHT (3 cycles, 30 minutes per cycle at 110 kHz and 30 kA/m). The UV-VIS adsorption and PL (wavelength of excitation 480 nm) spectra were recorded on the RPMI media upon removal of the scaffolds. a) The absorption peak centered at 480 nm, typical of DOXO, was too low in the case of the of PCL-23_IONCs-DOXO scaffold exposed to the magnetic hyperthermia treatment suggesting a very low amount of DOXO released in the RPMI solution after the three cycle of hyperthermia (lower than 1.25 µg/ mL of DOXO). b) and c) Similarly, the PL spectra of RPMI exposed to PCL-23_IONCs-DOXO, after magnetic hyperthermia, does not show a measurable and significant DOXO signal that can match with the characteristic DOXO peak of the DOXO dissolved in pure RPMI media. It is likely that after the three cycles of MHT only a tiny fraction of the incorporated DOXO is released and this amount is too low to be detected by UV-Vis and PL spectra. Moreover, we cannot exclude that the residual polymer, released by the scaffolds upon exposure to the MHT treatment, does not interfere with the DOXO signal in the UV-Vis and PL spectra.

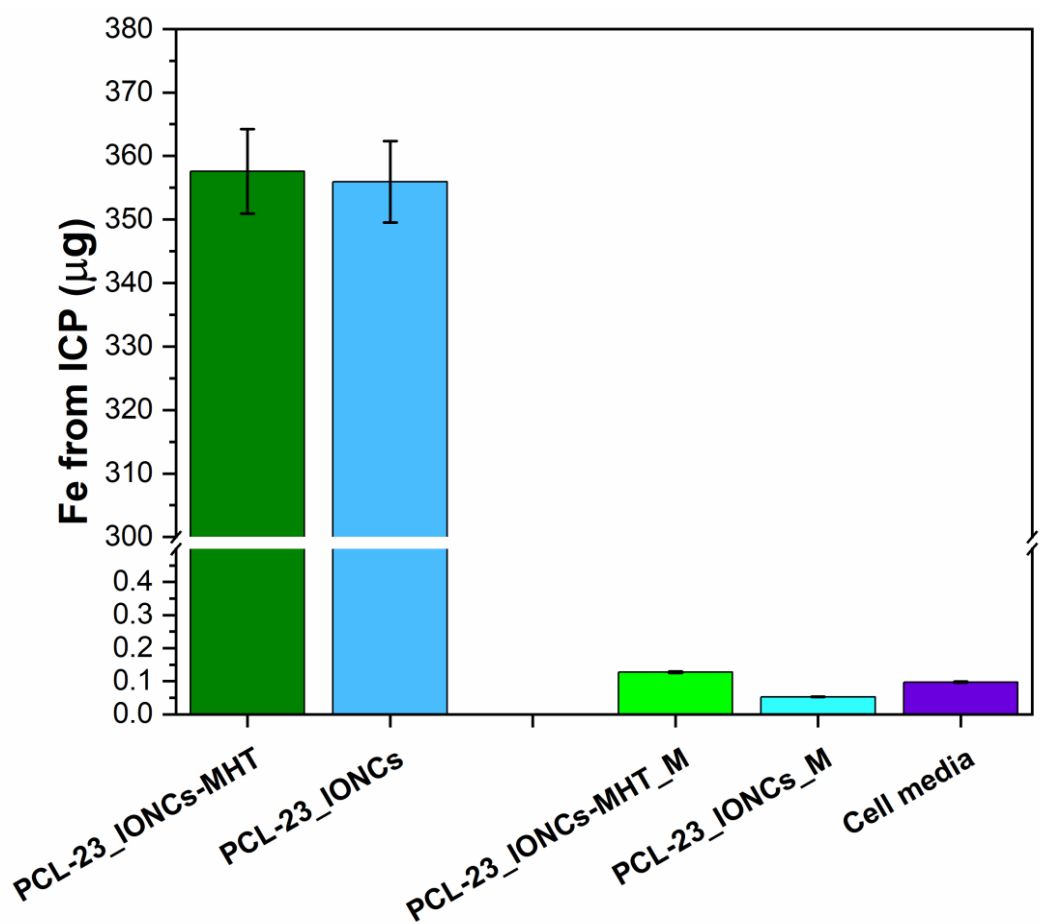

**Figure S2. Evaluation of the Fe release by elemental analysis using inductively coupled plasma optical emission spectroscopy (ICP-OES) after magnetic hyperthermia treatment.** The PCL-23_IONCS-MHT in phenol-free RPMI media samples were exposed to MHT treatment (three cycles, 30 minutes for each cycle at f= 110 kHz and T=45°C obtained by H varying from 24 to 30 kA/m adjusted automatically by the AMF applicator to maintain the temperature:). The control samples (PCL-23_IONCs in phenol-free RPMI media), was not exposed to MHT but in this case the scaffold was left 90 minutes at room temperature in RPMI media. The iron content was measured by elemental analysis, separately, for the scaffolds and for their relative media: PCL-23_IONCs-MHT

scaffold (dark green) and its media (PCL-23_IONCs-MHT_M in light green), PCL-23_IONCs scaffold (in blue) and its media (PCL-23_IONCs_M in light blue). A sample of just phenol-free RPMI media (Cell media) was used as a control and tested for the iron quantification of the blank (in violet).

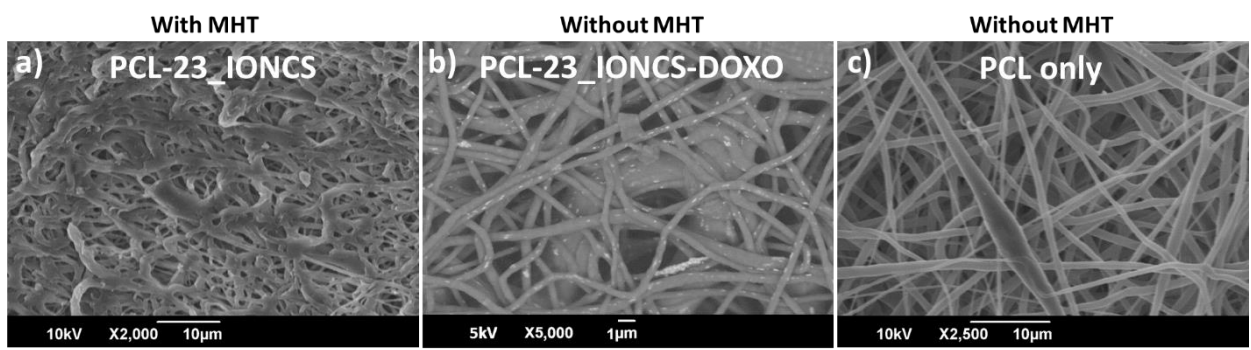

**Figure S3. Scanning electron microscopy (SEM) characterization of magnetic PCL scaffolds with and without MHT.** a) SEM images of the PCL-23_IONCs after three cycle (30 min each) of MHT. The MHT was performed at frequency of 110 kHz and field amplitude was adjusted automatically (from 24 to 30 kA/m) by the AMF applicator to maintain the RPMI solution temperature of 45°C. b and c) SEM images of PCL-23_IONCs-DOXO and PCL only scaffolds after being immersed in RPMI media for 90 minutes and with no exposure to MHT. A clear structural modification of the fibers occurs after MHT with respect to fibers not exposed to MHT.

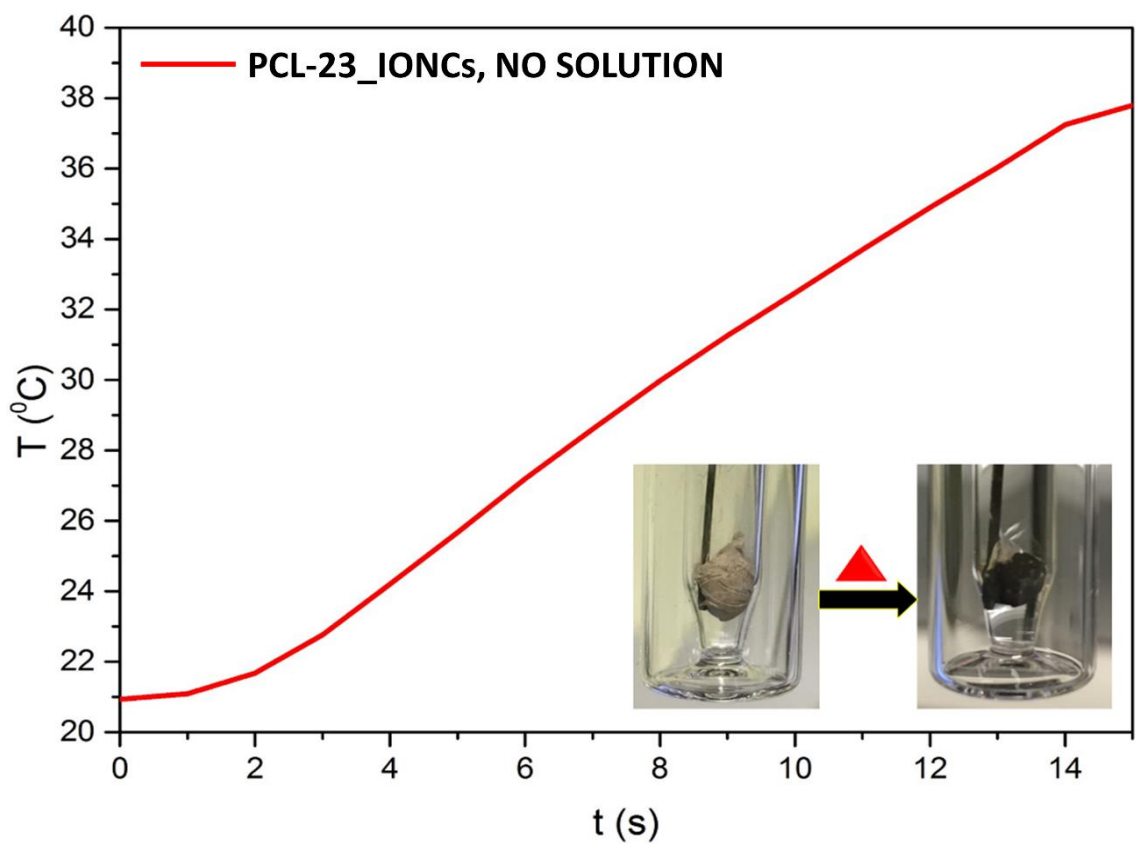

**Figure S4. MHT heating profile of PCL-23_IONCs in absence of solution:** when recording with an optical fiber placed in contact with the scaffold, the temperature increases upon application of alternating magnetic field (*f:110 kHz and H: 30 kA/m*), a rapid increment of temperature (ca. 18 °C) was reached just after 14 seconds of field applications. The insets, are photos of the scaffold before and after the heating cycle (120 seconds) showing how the scaffold shrinks and changes its appearance.

Published in Journal of Colloid and Interface Science, Volume 607, Part 1, February 2022, Pages 34-44, Doi: https://doi.org/10.1016/j.jcis.2021.08.153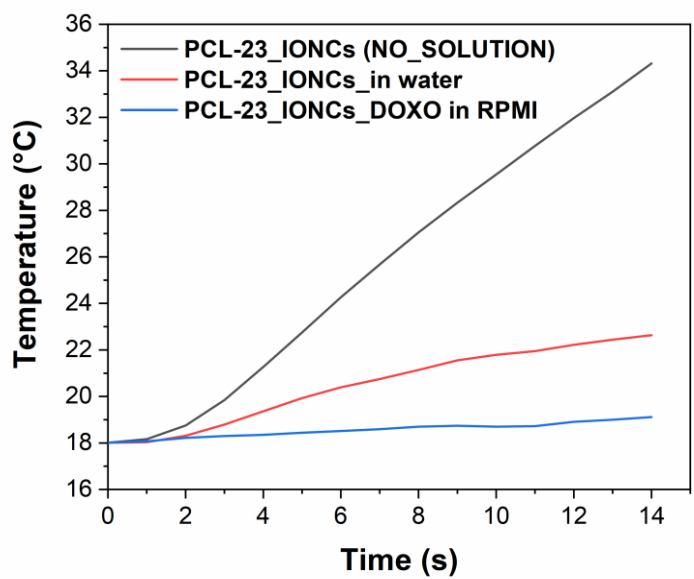

**Figure S5.** Temperature versus time curves of the first 14 seconds after having switched on the MHT (110kHz and 30 kA/m) for the same fibers measured in air (black curve), in water (red curve) and in RPMI media (blue curve).

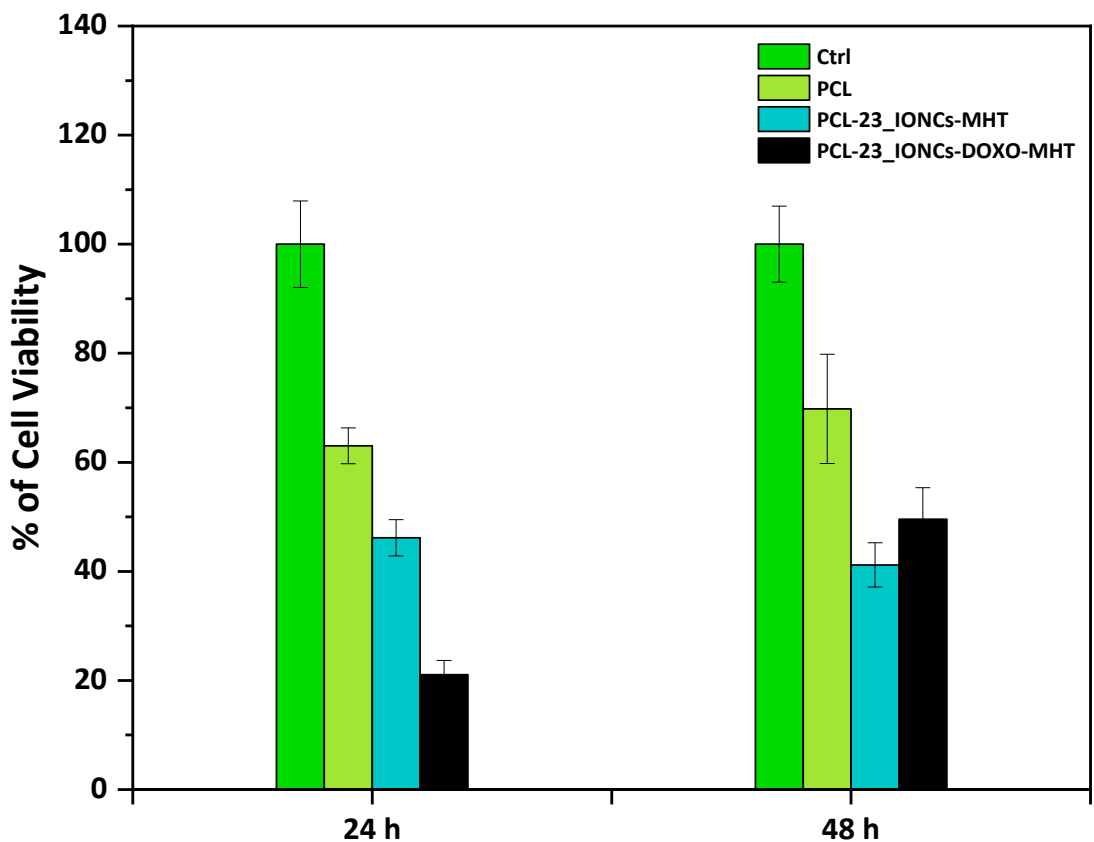

**Figure S6. MCF7 cells viability study after exposure to PCL scaffold loaded with IONCs in response to MHT treatment alone and MHT combined with DOXO release**. The control (CTRL, in green) represents MCF7 cell line alone. As can be seen from the chart, PCL scaffold alone showed mild toxicity (in light green), possibly caused by the non-sterilization of the scaffolds. The samples with MHT treatment alone (PCL-IONCs-MHT, in light blue) and the one with combined treatment (PCLIONCs-DOXO-MHT, in black) show a reduction of cells proliferation. Interestingly, there is an acute toxicity of the DOXO after 24h from the treatment, followed by a recovery in the following day. The thermal treatment seems slightly more effective than the one combined with DOXO. The MHT treated samples were exposed to three cycles (30 minutes each) of AMF (f=110 kHz H= from 24 to 30 kA/m adjusted automatically buy the AMF applicator to maintain the temperature: T=45°C). The viability was assessed by MTT assay at 24h, 48h from the treatment.

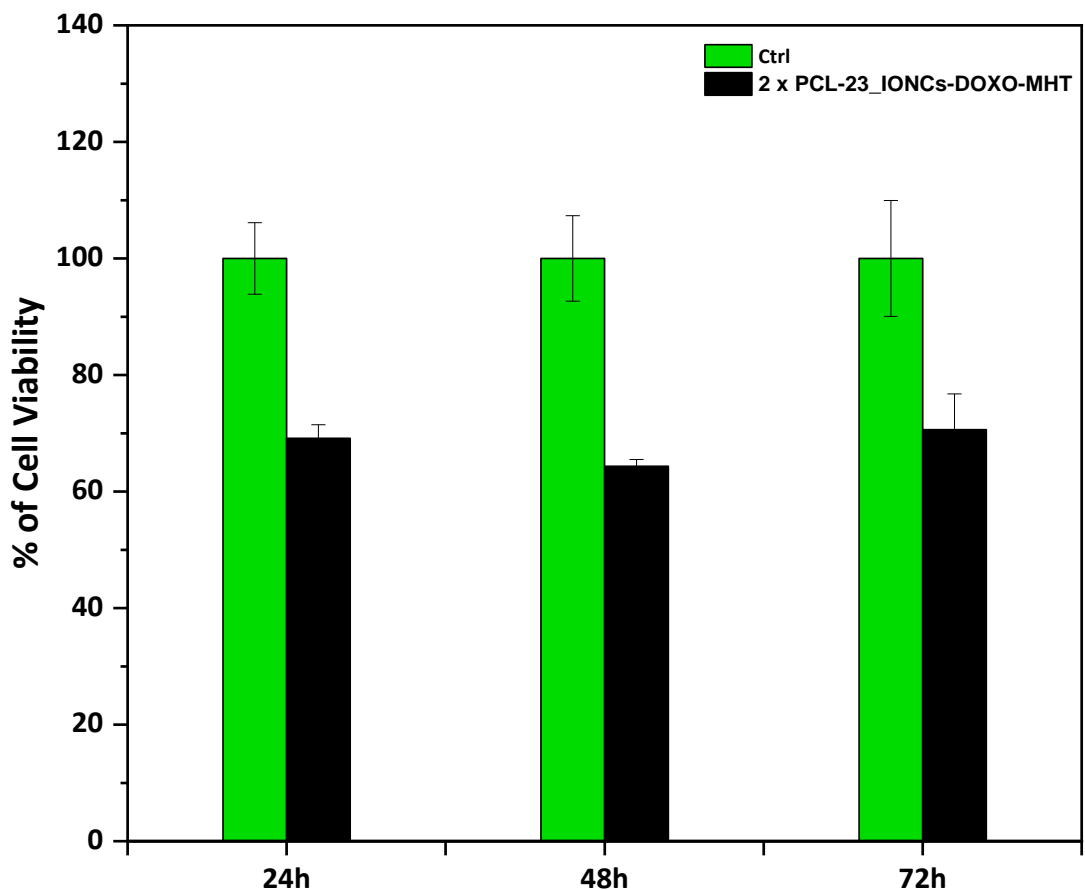

**Figure S7. MCF7 cells viability study after exposure to two PCL scaffold loaded with IONCS and DOXO**. The control (in green) represents MCF7 cell line alone, without the application of any treatment. As can be seen from the chart, for the cells exposed to two scaffolds (2xPCL-IONCs-DOXO-MHT, in black) and to MHT there is not the presence of acute toxicity after 24h from the treatment. The viability slightly reduces at the 48h (around 60%) and then it increases again at 72h after the treatment. The MHT treated samples were exposed to three cycles (30 minutes each) of AMF (f=110 kHz H= from 24 to 30 kA/m adjusted automatically buy the AMF applicator to maintain the temperature: T=45°C). The viability was assessed by MTT assay at 24h, 48h and 72h from the treatment.